\documentstyle[12pt]{article}
\begin{document}
\begin{center} 
{\bf Billiard Sequences and the Property of Splittability of
Integrable Hamilton Systems}\\[0.4cm]
A.~Yu.~Shahverdian \\ [0.2cm]
\begin{small}
(Yerevan Physics Institute) \\ [0.8cm]
\end{small}
\end{center} 
{\bf 1.~~Introduction} \\[0.2cm]
There are two kinds of billiards in $(n+1)$ dimensional unit cube
$D^{n+1}$, $(n\geq 1)$, differing by accepted reflection law, are
considered in the classical mechanics and dynamical systems theory.
It is the usual billiards with the reflection law -- the angle of
incidence is equal to angle of reflection, and the bouncing
billiards, when a particle, moving whithin cube rectilinearly
and uniformly in a given direction $\vec{\theta} =
(\theta^{1},\ldots, \theta^{n+1})$, if reaching any edge of cube
is bounced
vertically to the opposite one and then goes parallelly to its previous
path. On the other hand, there are known two different kinds of
one dimensional periodical motion: the rotation and libration [1-4].
In the case of bouncing billiards each of the
coordinates of the orbit's point performs the rotation,
whereas in the case of usual billiards each of them performs
the libration; in either cases the frequencies are equal to
components of vector $\vec{\theta}$.
The orbits of both motions, if considering
on the $(n+1)$ dimensional torus $Tor^{(n+1)}$ and cube $D^{n+1}$
respectively, are the continuous lines
which either are closed (periodical motion) or, under assumption
that $\vec{\theta}$ is nonresonance vector, according to H.~Weyl
theorem ([5], see also [2, 3, 6]), are distributed uniformly
on the whole surface of
$Tor^{(n+1)}$ or in cube $D^{n+1}$ (conditionally periodic motion).
For instance, the Kepler
problem on motion of planets [2] can be considered in terms of
bouncing billiards, while  for
the problems on small vibrations of
the pendulum [1, 2], the Lissazhu figures [1, 2], and the
vibrations of atom [1] it is
natural to consider the orbits are given in the cube,
in cartesian coordinates, and not on the torus.

We obtain the terms ${\vec x}_{k} \in D^{n}$ of the
($l$-th, $1\leq l \leq n+1$)
billiard sequence from the consecuitive points
${\vec z}_{k}$ of intersections of an usual billiard's
orbit with two opposite edges of $D^{n+1}$, if erase
from every vector $\vec{z}_{k}$ the ($l$-th)
component which is equal to $0$ or $1$.
The main result of present paper
establishes that under some simple
assumption on the reduced frequency $\theta^{i+1}/\theta^{l}$,
the ($i$-th, $1\leq i \leq n$) by-coordinate billiard
sequence $(x_{k}^{(i)})_{k=1}^{\infty}$ can
be splitted into countable
set of countable pairwise disjoint subsequences
in such a way that the restriction of the return map
$\pi_{i}: x_{k}^{(i)}\to  x_{k+1}^{(i)}$
on each of them appears the rotation transformation of the form
\begin{equation}
  a + b \{ \alpha + \beta k\} \qquad (k=1, 2, \ldots )
\end{equation}
of the circle $\{x: |x-a|=|b| \}$.
Such kind of splitting into finite number of
subsequences with the simultaneous exact estimating the density
of arising exceptional sets is also considered. We
note that the return map ${\vec \pi} =
(\pi_{1}, \ldots, \pi_{n})$ is similar
to Poincare map, which inherits [4] the basic features of the whole orbit.

It is known that the libration occurs in the
neighborhood of the elliptic points in phase space [1-4, 7].
Hence, in spite of our Theorems 1-4 relate to
billiards in cube, we note that through use of canonical
transformations [1-3], they are also applicable to the
integrable and weak perturbated Hamilton systems.
By this reason, the corresponding theoretical-mechanical
reformulations -- the Corollaries 1-3, are also provided.

Note in the end, that due to the results below, the
coexistence in weak chaotic Hamilton systems
of two (undistinguished in small bins of phase space) different
kinds of motion, mentioned by I.~Prigogine (for the case of weak
stable systems, [7], Ch.~2), is reduced to known coexistence
of the discrete and continuous forms. \\[0.3cm]
{\bf 2.~~Main results}\\[0.2cm]
This work contains improved formulations of
our results on billiard boundary sequences, announced before in
the note [8]; the corresponding theoretical-mechanical
reformulations are also given.
In order to emphasize the underlying dynamical context, in
the latter statements the terminology "conditionally periodic
libration in cube" is used.
In the "action-angle" coordinates [1-4, 9-11],
if we consider the "angle" variables as
the cartesian coordinates, the orbits of such motion
are exactly the usual billiard's orbits.

As in [8], we define
the billiard sequence of usual billiard in cube
$D^{n+1}$ as follows. Let for a pregiven index
$1\leq l\leq n+1$ the ${\vec z}_{k}$ be all the points
of an orbit for each of which
$z_{k}^{(l)} = 0, 1$ (it is supposed that
the orbits do not intersect the ribs of cube
$D^{n+1}$ and the index $k\geq 1$ increases with time).
The sequence
$B_{\vec \theta} = ({\vec x}_{k})^{\infty}_{k=1}$
of those points from
$D^{n}$, for which $x_{k}^{(s)}=z_{k}^{(s)}$ whenever
$1\leq s\leq l-1$ and
$x_{k}^{(s)}=z_{k}^{(s+1)}$ whenever  $l < s \leq n$
we call the ($l$-th) billiard (boundary) sequence.
We note that the Poincare section of
bouncing billiard's orbit consists of the
points of the form (see [4]),
\begin{equation}
 (\{ \alpha_{1}+\beta_{1}k \}, \ldots , \{ \alpha_{n}
   +\beta_{n}k \}) \qquad (k=1,2, \ldots)
\end{equation}
($\{.\}$ is the fractional part of number). It is not
difficult to show,
for usual billiards their billiard sequences $B_{\vec{\theta}}$
are of the form
\begin{equation}
(||\alpha_{1}+\beta_{1}k||, \ldots, ||\alpha_{n}+\beta_{n}k||)
\qquad (k=1,2,\ldots)
\end{equation}
where the quantity
  $$ ||x|| = 2\min(\{ x\},1-\{ x \}) $$
is so-called "the distance to the nearest integer"
of point $x\in R^{1}$, that is used in the theory of
Diophantine approximations [6, 12, 13].
In contrast to sequences of fractional parts from Eq.~(2), which
are studied quite well (see details in [6]), the properties
of sequences (3) are almost unknown (unless the results from [8],
[14] and several interesting but isolated number-theoretical
relations, can be found in [6], Ch.~2).

Our results relate to structure and properties of the
by-coordinate sequences
$B_{\vec{\theta},i} = (x_{k}^{(i)})_{k=1}^{\infty}$ and their
return map $\pi_{i}: x_{k}^{(i)} \to x_{k+1}^{(i)}$ under two
assumptions,
which are mentioned in all further formulations:
\begin{equation}
   1/2 < \omega_{i} < 1  \quad \mbox{where} \quad
   \omega_{i} = |\theta^{(i+1)}/\theta^{(l)}|
   \qquad (1\leq i\leq n) \ ,
\end{equation}
and the (reduced) frequency $\omega_{i}$ is irrational number.
The restriction (4) (which is equivalent to such one, accepted
in [8]) on the billiards under consideration
is arised from  the method applied.
This condition appears to be very natural and is
quite enough (due to some property of symmetry of
the discrepancy) in order to research the discrepancies
of the general sequences of fractional parts ([14]).

The next Theorem 1 establishes the property of splittability of
billiard by-coordinate
sequences $B_{\vec{\theta},i}$ into countable number of
subsequences of fractional parts of form (1).
Theorem 2 estimates the rate of the split process in dependence on
arithmetical characteristics of the frequency $\omega_{i}$.
The metrical Theorem 3 and the Theorem 4 estimate this rate.
Also, we formulate the theoretical-mechanical analogies of
these theorems, the Corollaries 1-3,
are based on representation of by-coordinate
return map $\pi_{i}$ in a form of (countable or finite) set of
the discrete rotation transformations of circles.
\newtheorem{guess}{Theorem}
\begin{guess}
If the condition (4) is satisfied, then
there exists a partition of the natural series $N$
into countable number of countable and pairwise disjoint
subsets $N_{i,s}$,
\begin{equation}
N=\sum_{s=1}^{\infty}N_{i,s} \ , \
\qquad N_{i,s_{1}}\cap N_{i,s_{2}}= 0 \qquad (s_{1}\neq s_{2}) \ ,
\end{equation}
such that for some numbers 
$a_{i,s}, \  b_{i,s}$, \  $\alpha_{i,s}, \  \beta_{i,s} $  
and every $s \geq 1$ the equalities
\begin{equation}
 x_{\nu_{i,s}(k)}^{(i)} =
a_{i,s}+b_{i,s}\{ \alpha_{i,s} + \beta_{i,s} \ k \}
\qquad (k=1, 2, \ldots)
\end{equation}
hold. Here the function $\nu_{i,s}$ is numbered the set  
$N_{i,s}$ in increasing order of its terms 
and each $\beta_{i,s}$
is irrational of the form $\beta_{i,s} = r_{i,s}({\omega}_{i})$
where $r_{i,s}$ is a fractional-linear transformation with entire 
coefficients.  
\end{guess} 
What is more, it can be also shown, that for each of the coefficients
$\beta_{i,s}$ from Eq.~(6), one of the following 5 simple
relationships on equivalence,
$$\beta_{i,s} \sim \omega_{i},
\quad \beta_{i,s} \sim \omega_{i}/2,
\quad \beta_{i,s}/2 \sim \omega_{i},
\quad 2\beta_{i,s}  \sim \omega_{i},
\quad 2\beta_{i,s} \sim 1- \omega_{i} $$
holds. We recall (see [12, 13]), that for large enough values of
index $k$ the $k$-th coefficients of expansion into
regular continued fraction of two equivalent
numbers are coincided.

In the formulations of
Corollaries 1-3, we use the terms
"rate of splitting"  and "boundary motion". The first one
means the density of the corresponding
exceptional set, while the latter means the
motion of terms of
the by-coordinate billiard sequences
$B_{{\vec \theta}, i} = x_{k}^{(i)}$,
determined by return map
\begin{equation}
\vec{\pi}=(\pi_{1}, \ldots, \pi_{n}) \quad
\mbox{and} \quad \pi_{i}(x_{k}^{(i)}) = \pi_{i}(x_{k+1}^{(i)})
\qquad (1\leq i \leq n; \  k=1,2,\ldots) \ .
\end{equation}
Indeed, if we introduce a discrete time
$t^{\prime}=\{ 1, 2, \ldots \} $
in such a way that the fall of moving particle on two opposite
edges of $D^{n+1}$ occures at its consecuitive moments ($k$ and
$k+1$),
the Eq.~(7) can be understood as the equation of motion
of the point ${x}_{k}^{(i)}$ in respect of $t^{\prime}$.
Furthermore, through the function $\nu_{i,s}$ defined in
Theorem 1, we
can introduce the splitting of the time series $t^{\prime}$ into
countable number of discrete independent times
$t^{\prime}_{i,s} (= N)$,
$$k\in t_{i,s}^{\prime} \quad <=> \quad
\nu_{i,s}(k) \in N_{i,s} \ . $$
Then, according to Theorem 1, the motion of point
$x_{\nu_{i,s}(k)}^{(i)}$
in respect of time $t_{i,s}^{\prime}=\{k:k\geq 1\}$ appears a
uniform discrete rotation of the circle
$\{ x:|x-a_{i,s}|=|b_{i,s}|\}$ with the frequency $\beta_{i,s}$.
The function $\nu_{i,s}$ can also be explicitly
expressed by some characteristics (the
functions $\lambda$ and $\tau$ below)
of the frequency $\omega_{i}$.
\newtheorem{guess2}{Corollary}
\begin{guess2}
Let us have a conditionally periodic libration in cube $D^{n+1}$
and for some $1\leq i \leq n$ the condition (4) is satisfied.
Then the boundary motion of $i$-th coordinate $x_{k}^{(i)}$ of
point $\vec{x}_{k}\in B_{\vec \theta}$ can be splitted
into a countable set of discrete uniform rotations.
\end{guess2}
In order to formulate the next two theorems we need some
definitions. For numbers $x\in (0,1/2)$ we consider the
expansion into continued fraction of the form (see [8]):
\begin{equation}
               x = \frac{\epsilon_{0}|}{|a_{1}}+
                   \frac{\epsilon_{1}|}{|a_{2}} + \cdots +
                   \frac{\epsilon_{k}|}{|a_{k+1}} + \cdots              
\end{equation}                         
where $\epsilon_{0}=1$ and for $k\geq 1$ the
coefficients $a_{k}=a_{k}(x)$ and
$\epsilon_{k}=\epsilon_{k}(x)$ are defined as follows  
\[a_{k} = \left\{ \begin{array}{ll}
   \:   [1/\psi^{k-1}]   &  \ \mbox {$\{ 1/\psi^{k-1}\}\leq 1/2$}\\  \:
        [1/\psi^{k-1}]+1 &  \ \mbox {$\{ 1/\psi^{k-1}\}>1/2$}  
                \end{array}  \right. , \      
   \epsilon_{k} =\left\{ \begin{array}{ll}
                         $ 1$ & \  \mbox{$\{ 1/\psi^{k-1}\}\leq 1/2$}\\
                         $-1$ & \  \mbox{$\{ 1/\psi^{k-1}\}> 1/2$} \ . 
                        \end{array}  \right. \]   
Here $[.]$ and $\{.\}$ are the entire and fractional parts 
of number respectively, $\psi^{0}(x)\equiv x$, and for 
$k\geq 1$ \ $\psi^{k}$ is $k$-th iterate of the function $\psi$,
\begin{equation}
 \psi (x) = \min \ (\{1/x\}, 1-\{1/x\}) \quad (= \frac{1}{2}||1/x||) \ .
\end{equation}
We use also the binary coefficients $\rho_{k}$: $\rho_{k}=0$ whenever
$\epsilon_{k}=-1$ and  $\rho_{k}=1$ whenever 
$\epsilon_{k}=1$
and rewrite the expansion (8) in the form
\begin{equation}
 x = [a_{1},a_{2},\ldots ; \rho_{1}, \rho_{2},\ldots] \ .
\end{equation}
This modified expansion appears to be more appropriate
tool for considering 
problems (as well as for problems concerning the discrepancy, see
[14]), than the regular continued fractions.
There is the next interrelation between the coefficients
$a_{k}(x)$ and $\rho_{k}(x)$ and the coefficients  
$b_{k}(x)$ of expansion into regular
continued fraction of number $x$ (see [8]):
\begin{equation}
 a_{k}= b_{\sigma(k)}+2-\rho_{k} - \rho_{k-1}, \qquad \rho_{k}=0  \  < = >  
\  b_{\sigma(k)+1} =1   
\end{equation}
where the quantity $\sigma \ (=\sigma_{x})$ 
is defined as follows:   
$\sigma(1)=1$, \  $\sigma (k+1) = 
\sigma (k) + 2 - \rho_{k}$ \ $(k \geq 1)$. 
We define  also the functions
$\lambda \ (= \lambda_{x})$ and $\tau \ (=\tau_{x})$, are mentioned
in our formulations: $\lambda (1)=1,
\ \lambda (k+1)=\lambda (k)+\tau (k)+1$ where 
$\tau (1) = 1-\rho_{1}$ and $\tau (k+1)=\tau (k)$ 
whenever $a_{\lambda (k+1)}$ is odd, 
$\tau (k+1)=1-\tau (k)$ whenever  
$a_{\lambda (k+1)}$ is even.  
For a set
$E\subset N$ its density is defined as
$$dens (E) = \lim _{p\to \infty}p^{-1}|\{k\in E:1\leq k\leq p\}|)
\quad \mbox{(here $|.|$ is power of the set)} \ . $$
The Theorems 2 - 4 express
the rate of the split process, established in Theorem 1 (and
which, we have assumed, is measured by the density of
exceptional set), by the number-theoretical characteristics
of the reduced frequency $\omega_{i}$.
\begin{guess} 
If the condition (4) is satisfied, and the function
$\mu \ (= \mu_{i})$ is
\begin{equation}
\mu (p) = 2p +
2\sum_{k=1, \tau (k)=1}^{p} \rho_{\lambda (k)+1}
\qquad (p=1, 2, \ldots)
\end{equation}
where $\lambda = \lambda_{\omega_{i}^{\prime}}\ $,
$\tau=\tau_{\omega_{i}^{\prime}}\ $, and
$\omega_{i}^{\prime} = 1-\omega_{i}$ ,
then for every $p\geq 1$ there is a finite partition
of natural series into pairwise disjoint subsets,
\begin{equation}
  N=\sum_{s=1}^{\mu(p)}N_{i,s} + E_{i,p}
\end{equation}
such that for each $1\leq s\leq \mu$
and all $k\geq 1$ the equations (6) hold and
for the density of exceptional set 
$E_{i,p}$ we have   
\begin{equation}
   dens(E_{i,p}) = 
\prod_{s=0}^{\lambda (p+1)-1}\psi^{s}(\omega_{i}^{\prime})
= \prod_{s=1}^{\lambda (p+1)}[a_{s}(\omega_{i}^{\prime}),
\ldots; \rho_{s}(\omega_{i}^{\prime}), \ldots ] \ .
\end{equation}
\end{guess} 
By virtue of the relationships (11), the value
of $dens(E_{i,p})$ in (14) can also be
expressed directly 
through the coefficients $b_{k}$ of the frequency $\omega_{i}$.
In such a way, we have all the quantities
$\lambda$, $\mu$ and
$dens(E_{i,p})$ in explicit form. Hence, it is not
difficult to obtain from Theorem 2 many interesting
particular relations giving the dependence of density of 
exceptional set $E_{i,p}$  
on the arithmetical properties of the coefficients
$a_{k}$ and  $\rho_{k}$ or $b_{k}$. 

The next statement is a reformulation of Theorem 2.
\begin{guess2}
Let us have a conditionally periodic motion in cube $D^{n+1}$ and
for some $1\leq i\leq n$ the condition (4) holds.
Then for every $p\geq 1$ the boundary motion of $i$-th
coordinate $x_{k}^{(i)}$ of the point
$\vec{x}_{k}\in B_{\vec \theta}$ can
be splitted into a finite set of $\mu_{i}(p)$
discrete uniform rotations and the split rate is equal to
right-hand expression in Eq.~(14).
\end{guess2}
\begin{guess}
If the condition (4) is satisfied, then for a.e. (in sense
of Lebesgue measure) numbers $\omega_{i}\in (1/2, 1)$ and
enough large $p\geq 1$
the sequence $B_{\bar{\theta}, i}$
can be splitted into
$\mu_{i}$, $3p\leq \mu_{i} \leq 4p$ subsequences of
fractional parts of the form (6) in such a way,
that for the density
of exceptional set $E_{i,p}$ we have
$$ dens(E_{i,p}) = (1+o(1))\exp \: (- \gamma \: p) 
\qquad (p\to \infty)$$
where  $0< \gamma < \infty$ is an absolute constant.   
\end{guess}  
By use of the explicit form of
invariant measure of the ergodic shift
$\psi$,
\begin{equation}
\psi([a_{1}, a_{2}, \ldots; \rho_{1}, \rho_{2}, \ldots]) =
[a_{2}, a_{3},\ldots; \rho_{2}, \rho_{3},\ldots ] \ ,
\end{equation}
defined by Eq.~(9) (see [8]), one can obtain
the exact numerical value of this constant (see [8]):
 $$ \gamma =
 \frac{3}{2}\frac{Li_{2}(\frac{3-\sqrt{5}}{4}) -
 Li_{2}(\frac{1-\sqrt{5}}{4})}
 {\ln \frac{1+\sqrt{5}}{2}} = 1.524\ldots $$
where $Li_{2}$ is the Euler dilogarithm ([15]).
It is interesting to note that  such a constant, is
equal to $\frac{2}{3}\gamma$, is arised in a metrical theorem on
estimating the discrepancy (see [14]).

The point (b) of the next theorem follows from the previous
Theorem 3 and is given again just for completness.
\begin{guess}
If the condition (4) is satisfied, then
the sequence $B_{\vec{\theta},i}$ can be splitted into
$\mu$, $2p\leq \mu \leq 4p$ subsequences of
the form (6) in such a way, that for the density of the
exceptional set $E_{i,p}$ from Eq.~(13), the next
statements are true:\\
(a)~~there exists a constant $C > 0$ such that for every
$\omega_{i} \in (1/2, 1)$ we have
  $$ dens (E_{i,p}) = O(\exp(- C p))  \qquad (p\to \infty)\ ;$$
(b)~~for almost each $\omega_{i}\in (1/2, 1)$ we have
  $$ dens (E_{i,p}) = (1+o(1))\exp(-\gamma p) \qquad (p\to
  \infty) \ ; $$
(c)~~if the $\omega_{i} \in (1/2, 1)$ is a
Liouville number, then we have
  $$ dens (E_{i,p}) = o(\exp(- C(p) p)) \qquad (p\to \infty) $$
where $C(p)\to +\infty$ as $p\to \infty$; in dependens 
on $\omega_{i}$ the quantity $C(p)$ can tend to $+\infty$
as quickly as we like.
\end{guess}
The next corollary is the reformulation of Theorem 4.
The distance between two motions, as it is mentioned above,
should be understood in sense of density of the exceptional sets
$E_{i,p}$ from Eq.~(13).
\begin{guess2}
If the condition (4) is satisfied, then the
next 3 statements are true:~(a)~the (boundary motion of)
conditionally periodic libration
is ever approximated by a finite set of discrete rotations
at least with the exponential rate;~(b)~the
(boundary motion of) almost each conditionally periodic libration is
approximated by a finite set of
discrete rotations with the exponential rate;~(c)~there
exist the conditionally periodic librations (the boundary
motions of) which
are as close to a finite set of discrete rotations, as we like.
\end{guess2}
 
\end{document}